\documentclass[prb,fleqn,12pt,showpacs]{revtex4}
\usepackage{epsfig}
\usepackage{graphicx}
\usepackage{amsfonts}
\begin{document} 
\title{Fermionic representation \\ for the ferromagnetic Kondo lattice model --- \\ 
diagrammatic study of spin-charge coupling effects \\ on magnon excitations}
\author{Sudhakar Pandey, Subrat Das, Bhaskar Kamble, Saptarshi Ghosh, Dheeraj Singh, Rajyavardhan Ray, and Avinash Singh}
\email{avinas@iitk.ac.in} 
\affiliation{Department of Physics, Indian Institute of Technology Kanpur - 208016}
\begin{abstract}
A purely fermionic representation is introduced for the ferromagnetic Kondo lattice model which allows conventional diagrammatic tools to be employed to study correlation effects. Quantum $1/S$ corrections to magnon excitations are investigated using a systematic inverse-degeneracy expansion scheme which incorporates correlation effects in the form of self-energy and vertex corrections, while explicitly preserving the continuous spin-rotation symmetry. Magnon self-energy is studied in the full range of interaction strength, and shown to result in strong magnon damping and anomalous softening for zone boundary modes, which accounts for several zone-boundary anomalies observed in recent spin-wave measurements of ferromagnetic manganites.

\end{abstract}
\maketitle
\newpage
\section{Introduction}
The interplay between itinerant carriers in a partially filled band and localized magnetic moments is conventionally 
studied within the ferromagnetic Kondo lattice model (FKLM), wherein $- J {\bf S}_I . {\mbox{\boldmath $\sigma$}}_I $ 
represents the local exchange interaction between the localized magnetic moment ${\bf S}_I$ and the itinerant electron spin ${\mbox{\boldmath $\sigma$}}_I$. Magnon excitations in the FKLM provide important information about this interplay, with the magnon dispersion determining the finite-temperature spin dynamics, transition temperature $T_c$, and also providing information about the carrier-induced spin-spin interactions in the ferromagnetic state, negative-energy modes signalling instability due to competing antiferromagnetic (AF) interactions, and magnon damping highlighting the characteristic spin-charge coupling in a band ferromagnet.

Magnons in the FKLM have been studied as function of electron (hole) density $n$ ($p$) in the conduction (valence) band 
and the spin-fermion coupling $J$ in the context of heavy fermion materials,\cite{sigrist_92} ferromagnetic metals Gd, Tb, Dy, doped EuX,\cite{donath_98} colossal magnetoresistive manganites (CMR),\cite{furukawa_96,sarker_96,wang_98,vogt_2001} 
ordered diluted ferromagnets\cite{subrat_2005} relevant for the ordered double perovskite 
$\rm Sr_2Fe Mo O_6$,\cite{kobayashi_98} and diluted magnetic semiconductors such as $\rm Ga_{1-x} Mn_x As$ in which impurity positional disorder and clustering play a very significant role on magnon excitations and hence on the finite-temperature spin dynamics.\cite{asingh_2007,as_prb2007} 

Quantum corrections to the magnon spectrum at finite $S$, not accounted for in the above leading-order calculations, have also been studied using various approaches such as exact diagonalization,\cite{zang_97,kaplan_98} 
modified RKKY approach,\cite{nolting_97} variational method,\cite{okabe_98,wurth_98} the Holstein-Primakoff transformation.\cite{golosov_2000,shannon_2002} Most of these theoretical investigations\cite{Kapetanakis-PRB06} were carried out in the strong-coupling (double-exchange) limit ($J/W \gg 1$), where $W$ is the bandwidth of the mobile electrons and $J$ is their exchange coupling to the localized spins. Although providing a good description of the experimentally observed magnon damping in the ferromagnetic phase of colossal magneto-resistive (CMR) manganites,\cite{Zhang-JPCM-07}, these investigations, however, could not satisfactorily account for the characteristic anomalous magnon softening observed for zone-boundary modes.

For ferromagnetic manganties, typical parameter values are: $t=W/12 \sim 0.2-0.5$ eV for the mobile ($e_g$) electrons 
and $2J \sim 2$ eV for their exchange coupling to the localized core ($t_{2g}$) spins,\cite{dagotto_2001} 
so that with $2J/t \sim 4 - 10$, the intermediate-coupling regime ($J/W \sim 1$) appears to be more appropriate.
Theoretical studies have therefore been carried out recently in this regime as well, such as the variational investigation,\cite{Kapetanakis-PRB06} where anomalous softening is pronounced for $J/W \sim 1$. Furthermore, by taking into account the Coulomb repulsion between band electrons, which is the largest energy scale in manganites and often omitted in conventional FKLM investigations, several realistic features such as doping dependent asymmetry of the ferromagnetic phase and enhanced zone-boundary anomalous softening have been demonstrated, thereby highlighting the importance of correlated motion of electrons on spin dynamics.\cite{Golosov-PRB05,Kapetanakis-PRB07}  

Spin dynamics in the ferromagnetic phase of CMR manganites has attracted considerable current interest.\cite{Zhang-JPCM-07} Recent spin-wave excitation measurements have revealed several anomalous features in the magnon spectrum near the Brillouin-zone boundary.\cite{Ye-PRL06,Ye-PRB07,Moussa-PRB07} These observations are of the crucial importance for a quantitative understanding of the carrier-induced spin-spin interactions and magnon damping, 
and have highlighted the limitations of various existing theoretical approaches. 
For example, the predictions of disorder-induced softening\cite{Furukawa-PRB05} 
of the zone-boundary modes and magnon-phonon coupling\cite{Dai-PRB00} as the origin of magnon 
damping are in sharp contradiction with these observations.
Furthermore, the dramatic difference in the sensitivity of long-wavelength and
zone-boundary magnon modes on the mobile charge carriers has emerged as one of the
most puzzling feature. Observed for a finite range of hole concentrations, while the
spin stiffness remains almost constant, the softening and broadening of the 
zone-boundary modes show substantial enhancement with increasing hole concentrations.\cite{Ye-PRL06,Ye-PRB07}

It is therefore of interest to develop an approach wherein quantum fluctuation effects associated with both finite $S$ and electron correlation can be studied on an equal footing. In this paper we will introduce a purely fermionic representation of the FKLM in terms of a multi-orbital Hubbard model, which allows conventional diagrammatic tools to be employed for the investigation of quantum corrections beyond the leading order. We will make use of the inverse-degeneracy expansion scheme, employed recently to investigate ferromagnetism in the Hubbard model,\cite{as+sp} which systematically incorporates correlation effects in the form of self-energy and vertex corrections, while preserving the spin-rotational symmetry and hence the Goldstone mode {\em explicitly}. This purely fermionic analysis can be seamlessly extended to correlated band electrons, which will be reported separately.  

The organization of this paper is as follows. A purely fermionic representation for the FKLM is introduced in section II.
The magnon propagator is discussed in section III in terms of the irreducible particle-hole propagator, for which the first-order quantum corrections incorporating self-energy and vertex corrections are obtained in section IV within a systematic inverse-degeneracy expansion scheme. Results for the magnon self energy, renormalized magnon energy and damping, and electronic spectral-weight transfer due to electron-magnon coupling are presented in section V for one-, two-, and three-dimensional cases. Finally, conclusions are presented in section VI. 

\section{Two-orbital Hubbard model}
The conventional FKLM model involves mobile electrons in a partially-filled band exchange-coupled to localized spins. In this section we introduce a purely fermionic representation in terms of a multi-orbital Hubbard model, which allows conventional diagrammatic tools to be employed to investigate quantum corrections beyond the random phase approximation (RPA). 
For concreteness, we start with a two-orbital Hubbard model:
\begin{equation}
H=\sum_{i\sigma} \epsilon_\alpha a^\dagger _{i\alpha\sigma}  a_{i\alpha\sigma}
+ \sum_{{\bf k}\sigma} \epsilon_{\bf k} a^\dagger _{{\bf k}\beta\sigma}  a_{{\bf k}\beta\sigma}
- U \sum_i {\bf S}_{i\alpha}.{\bf S}_{i\alpha}
- 2J \sum_i {\bf S}_{i\alpha}.{\bf S}_{i\beta} 
\end{equation}
with a magnetic ($\alpha$) orbital and a non-magnetic ($\beta$) orbital.
Here the correlated and localized $\alpha$ orbital with large Hubbard interaction $U$
yields exchange-split spin bands with vanishing bandwidth and,
with occupancies $n_{\alpha \uparrow} = 1$ and $n_{\alpha \downarrow} = 0$ at half-filling,
represents spin $S=1/2$ localized magnetic moments.
Spin-$S$ magnetic moments, such as $S=5/2$ in DMS systems and $S=3/2$ in manganites,
can be similarly represented by simply including multiple (${\cal N}=2S$) $\alpha$ orbitals per site
with strong Hund's coupling. The $\beta$ band with dispersion $\epsilon_{\bf k}$ represents mobile fermions 
such as valence-band holes in DMS or $e_g$ electrons in manganites. 
The Hund's coupling $J$ between the two orbital spins ${\bf S}_{i\alpha}$ and ${\bf S}_{i\beta}$
represents the conventional exchange interaction between the localized spin and the mobile fermion spin,
as considered phenomenologically in the ferromagnetic Kondo lattice model.
A correlation term can also be included for the mobile ($\beta$) fermions. 

In the saturated ferromagnetic state, with ordering chosen in the $\hat{z}$ direction,
and spatially uniform magnetizations $m_\alpha =1$ and $m_\beta = m$ for the localized and mobile fermions,
corresponding to fully occupied $\alpha\uparrow$ band and partially occupied $\beta\uparrow$ band,
the interaction terms reduce to: 
\begin{equation}
H_{\rm int}^{\rm HF} = 
-\sum_{i\mu} \psi_{i\mu}^\dagger [\sigma.{\bf \Delta_\mu}] \psi_{i\mu}
\end{equation}
at the Hartree-Fock level, where $\psi_\mu$ represent the fermion field operators for the two orbitals $\mu=\alpha,\beta$,
and the corresponding exchange splittings are given by:
\begin{eqnarray}
2\Delta_\alpha &=& 2 (U\langle S_{i\alpha}^z \rangle + J\langle  S_{i\beta}^z \rangle ) = (U+Jm) \nonumber \\ 
2\Delta_\beta  &=& 2 J\langle S_{i\alpha}^z \rangle = J \; .
\end{eqnarray}

The bare antiparallel-spin particle-hole propagators for the two orbitals are given by:
\begin{eqnarray}
\chi^0 _\alpha (\omega) &=& 
\frac{1}{2\Delta_\alpha +\omega -i \eta} = 
\frac{1}{U + Jm +\omega -i \eta} \nonumber \\
\chi^0 _\beta ({\bf q},\omega) &=& \sum_{\bf k}
\frac{1}{\epsilon_{\bf k-q}^{\downarrow +} - \epsilon_{\bf k}^{\uparrow -} + \omega -i \eta} 
= \frac{m}{J + \omega -i \eta} \;\;\; ({\rm for} \;\; {\bf q}=0) \; ,
\end{eqnarray}
where $\epsilon_{\bf k}^\sigma = \epsilon_{\bf k} - \sigma \Delta_\beta$ 
are the exchange-split band energies, and superscripts $+$ $(-)$ refer to particle (hole) states above (below) the Fermi energy $\epsilon_{\rm F}$.

\begin{figure}
\vspace*{-20mm}
\hspace*{-0mm}
\psfig{figure=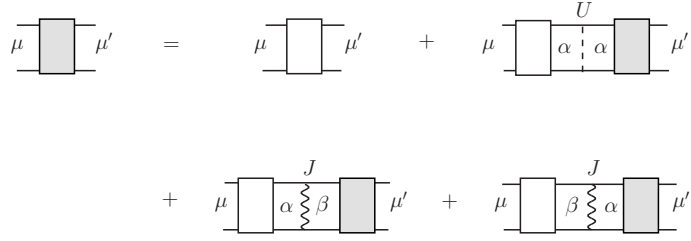,width=120mm}
\vspace{-120mm}
\caption{Diagrammatic representation of coupled equations for components of $\chi ^{-+}$ (shaded box) in terms of the irreducible particle-hole propagator 
$\phi({\bf q},\omega)$ (open box).}
\end{figure}

\section{Magnon propagator}
The different components $(\mu,\mu' = \alpha,\beta)$ of the 
time-ordered magnon propagator in the ferromagnetic state $| \Psi_0 \rangle$ are given by:  
\begin{equation}
\chi ^{-+} _{\mu\mu'} ({\bf q},\omega) =
i \int dt  \; e^{i\omega (t-t')} \sum_j e^{i{\bf q}.({\bf r}_i - {\bf r}_j)} 
\langle \Psi_0 |{\rm T}[S_{i\mu} ^- (t) S_{j\mu'} ^+ (t')]\Psi_0 \rangle 
\end{equation}
in terms of the spin-lowering and -raising operators $S_{i\mu} ^\mp = \psi_{i\mu} ^\dagger (\sigma^\mp/2) \psi_{i\mu}$.
As shown in Fig. 1, the coupled equations for the different components of $\chi ^{-+} $ can be expressed exactly as: 
\begin{eqnarray}
\chi ^{-+} _{\mu\mu'} &=& \phi_{\mu\mu'} + \phi_{\mu\alpha} U \chi ^{-+} _{\alpha\mu'}
+ \phi_{\mu\alpha} J \chi ^{-+} _{\beta\mu'} + \phi_{\mu\beta} J \chi ^{-+} _{\alpha\mu'} 
\end{eqnarray}
in terms of the irreducible particle-hole propagators $\phi$.
In analogy with the inverse-degeneracy $(1/\cal N)$ expansion studied recently for the band ferromagnet,\cite{as+sp}
we consider a systematic expansion:  
\begin{eqnarray}
\phi_{\mu\mu'} &=& \chi^0 _\mu \delta_{\mu\mu'} + \phi^{(1)}_{\mu\mu'} + \phi^{(2)}_{\mu\mu'} + ...
\end{eqnarray}
for the irreducible propagator, where $\chi^0$ are the bare particle-hole propagators, 
and correlation effects in the form of self-energy and vertex corrections are incorporated systematically in the quantum corrections $\phi^{(1)}$ etc. so that spin-rotation symmetry and hence the Goldstone mode are explicitly preserved order by order. 

Solving these coupled equations, we obtain:
\begin{eqnarray}
\chi ^{-+} _{\beta\beta} &=& \frac{\phi_{\beta\beta} (1 - U \phi_{\alpha\alpha}) + U\phi_{\beta\alpha} \phi_{\alpha\beta} } 
{1 - U\phi_{\alpha\alpha} - J^2 \phi_{\alpha\alpha} \phi_{\beta\beta} - J(\phi_{\beta\alpha} + \phi_{\alpha\beta}) 
+ J^2 \phi_{\alpha\beta} \phi_{\beta\alpha} } \\
\chi ^{-+} _{\beta\alpha} &=& \frac{\phi_{\beta\alpha} + J\phi_{\alpha\alpha} \phi_{\beta\beta} 
- J \phi_{\alpha\beta} \phi_{\beta\alpha} }
{1 - U\phi_{\alpha\alpha} - J^2 \phi_{\alpha\alpha} \phi_{\beta\beta} - J(\phi_{\beta\alpha} + \phi_{\alpha\beta}) 
+ J^2 \phi_{\alpha\beta} \phi_{\beta\alpha} } = \chi^{-+}_{\alpha\beta} \\
\chi ^{-+} _{\alpha\alpha} &=& \frac{\phi_{\alpha\alpha}}
{1 - U\phi_{\alpha\alpha} - J^2 \phi_{\alpha\alpha} \phi_{\beta\beta} - J(\phi_{\beta\alpha} + \phi_{\alpha\beta}) 
+ J^2 \phi_{\alpha\beta} \phi_{\beta\alpha} } \nonumber \\
&=& \frac{1}
{g_\alpha ^{-1} - J^2 \phi_{\beta\beta} - 2J\phi_{\alpha\beta} \phi_{\alpha\alpha}^{-1} 
+ J^2 \phi_{\alpha\beta}^2 \phi_{\alpha\alpha} ^{-1} }  
\end{eqnarray}
where 
\begin{equation}
g_\alpha (\omega) \equiv \frac{\phi_{\alpha\alpha}}{1-U \phi_{\alpha\alpha}}
\end{equation} 
represents the local magnon propagator for the localized spin ($\alpha$ orbital).
Resulting from the coupled nature of Eq. (6) for the different components, the common denominator in Eqs. (8-10) ensures a single Goldstone mode for all components, as expected. 

\begin{figure}
\vspace*{-20mm}
\hspace*{-0mm}
\psfig{figure=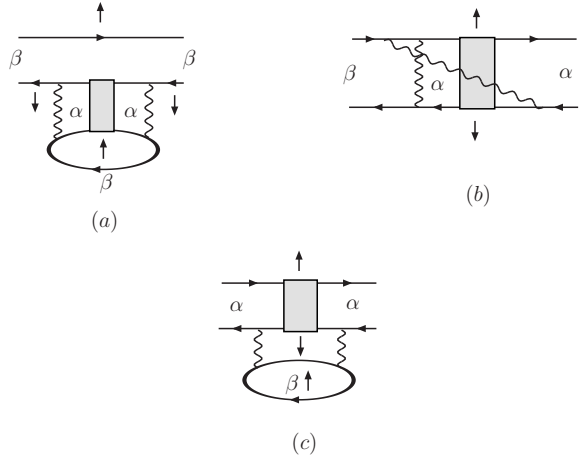,width=120mm}
\vspace{-100mm}
\caption{The first-order quantum corrections to the irreducible particle-hole 
propagator $\phi({\bf q},\omega)$.}
\end{figure}

The first-order quantum corrections to the irreducible propagator $\phi$ are shown diagrammatically in Fig. 2,
and correspond to the following physical processes: 
a) positive correction (to the bare particle-hole propagator) due to spin-$\downarrow$ spectral-weight transfer arising from self-energy correction, b) negative correction due to particle-particle correlations, and c) positive correction due to spin-$\downarrow$ correlations induced by spin-$\uparrow$ charge fluctuations. The corresponding expression are obtained as:
\begin{eqnarray}
\phi^{(1)}_{\alpha\alpha} &=& J^2 \left( \frac{1}{2\Delta_\alpha + \omega}\right )^2 
\sum_{\bf Q} \int \frac{d\Omega}{2\pi i} \; \chi^{-+}_{\rm RPA} ({\bf Q},\Omega)
\sum_{\bf k} \left ( \frac{1}
{\epsilon_{\bf k-q+Q}^{\uparrow +} -\epsilon_{\bf k}^{\uparrow -} +\omega -\Omega -i \eta} 
\right )
\nonumber \\
\phi^{(1)}_{\beta\beta} &=& J^2
\sum_{\bf Q} \int \frac{d\Omega}{2\pi i} \; \chi^{-+}_{\rm RPA} ({\bf Q},\Omega)
\sum_{\bf k} 
\left ( \frac{1}
{\epsilon_{\bf k-q}^{\downarrow +} -\epsilon_{\bf k}^{\uparrow -} +\omega -i \eta}
\right )^2 \left ( \frac{1}
{\epsilon_{\bf k-q+Q}^{\uparrow +} -\epsilon_{\bf k}^{\uparrow -} +\omega -\Omega -i \eta} 
\right )
\nonumber \\
\phi^{(1)}_{\alpha\beta} &=& -J^2 \left ( \frac{1}{2\Delta_\alpha + \omega} \right )
\sum_{\bf Q} \int \frac{d\Omega}{2\pi i} \; \chi^{-+}_{\rm RPA} ({\bf Q},\Omega) 
\sum_{\bf k} \left ( \frac{1}
{\epsilon_{\bf k-q}^{\downarrow +} -\epsilon_{\bf k}^{\uparrow -} +\omega -i \eta} \right )
\nonumber \\ 
& \times & \left ( \frac{1}
{\epsilon_{\bf k-q+Q}^{\uparrow +} -\epsilon_{\bf k}^{\uparrow -} +\omega -\Omega -i \eta} 
\right ) ,
\end{eqnarray}
where 
\begin{equation}
\chi^{-+}_{\rm RPA} ({\bf Q},\Omega) = \frac{\chi^0_\alpha(\Omega)}{1 - U\chi^0_\alpha(\Omega) 
- J^2 \chi^0_\alpha(\Omega) \chi^0_\beta({\bf Q},\Omega) } 
 \end{equation}
denotes the leading-order result for the magnon propagator $\chi ^{-+} _{\alpha\alpha}$ in the RPA. 

Separating out the bare part from the quantum corrections in the irreducible particle-hole propagators $\phi$ in (10), 
and dropping explicitly second-order terms such as $\phi^{(1)}_{\alpha\beta} \phi^{(1)}_{\beta\alpha}$, we obtain: 
\begin{equation} 
\chi ^{-+} _{\alpha\alpha}({\bf q},\omega) = \frac{1}
{[\omega + Jm - J^2 \chi^0 _\beta ({\bf q},\omega)] - 
[\Sigma_\alpha ({\bf q},\omega) + \Sigma_\beta ({\bf q},\omega)
+ 2\Sigma_{\alpha\beta} ({\bf q},\omega)]}
\end{equation}
for the $\alpha\alpha$ component.
The zeroth-order first term in the denominator yields the RPA result. 
The Goldstone mode at this level is easily verified, confirming the systematic nature of the expansion.
For ${\bf q}=0$, the bare fermion propagator $\chi^0 _\beta(0,\omega) = m/(J+\omega)$,
which yields a pole at $\omega =0$, as expected from the continuous spin-rotation symmetry.
For finite ${\bf q}$, the RPA magnon energy is obtained by solving the pole equation
\begin{equation}
\omega+Jm -J^2  \chi^0 _\beta({\bf q},\omega) = 0 .
\end{equation}
As $J^2 \chi^0 _\beta (0,0) = Jm$ from (4), 
and the $\omega$ dependence of the bare fermion propagator $\chi^0_\beta$ is relatively weak,
the RPA magnon energy is approximately obtained as: 
\begin{equation}
\omega_{\bf q} ^0 = J^2 (2S) [ \chi^0 _\beta(0) - \chi^0 _\beta({\bf q}) ] 
\end{equation}
for the spin-$S$ case. In the strong-coupling limit, and for dispersion $\epsilon_{\bf k} = -2t (\cos k_x +\cos k_y +\cos k_z)$ on a simple cubic lattice, the bare magnon energy $\omega_{\bf q} ^0 \approx (3t/S) \sum_{\bf k} n_{\bf k} \cos k_x (1-\gamma_{\bf q})$ is of order $1/S$ and has the Heisenberg form. 

The first-order magnon self energies in Eq. (12) are obtained as: 
\begin{eqnarray}
\Sigma_\alpha ({\bf q},\omega) &=& (U+Jm+\omega)^2 \phi_{\alpha\alpha} ^{(1)}({\bf q},\omega) \nonumber \\
\Sigma_\beta  ({\bf q},\omega) &=& J^2 \phi_{\beta\beta} ^{(1)}({\bf q},\omega) \nonumber \\
\Sigma_{\alpha\beta} ({\bf q},\omega) &=& J(U+Jm+\omega) \phi_{\beta\alpha} ^{(1)}({\bf q},\omega) ,
\end{eqnarray}
and we now discuss the resulting magnon renormalization. As the Goldstone mode is obtained at the RPA level itself, the magnon self-energy corrections must exactly cancel for $q,\omega=0$, as explicitly demonstrated below. The three magnon self energy terms, corresponding to the $\alpha$, $\beta$ and mixed contributions, can be combined in a simple expression:
\begin{eqnarray}
\Sigma_{\rm total} \equiv \Sigma_\alpha +\Sigma_\beta + 2\Sigma_{\alpha\beta} &=&
J^2 \sum_{\bf Q} \int \frac{d\Omega}{2\pi i} \; \chi^{-+}_{\rm RPA} ({\bf Q},\Omega)
\sum_{\bf k} \left [ \left ( \frac{1}
{\epsilon_{\bf k-q+Q}^{\uparrow +} -\epsilon_{\bf k}^{\uparrow -} +\omega -\Omega -i \eta} \right ) \right .
\nonumber \\
&\times & \left . \left (1-\frac{J}
{\epsilon_{\bf k-q}^{\downarrow +} -\epsilon_{\bf k}^{\uparrow -} +\omega -i \eta} \right )^2 \right ] ,
\end{eqnarray}
which identically vanishes for $q,\omega=0$.
Equation (18) highlights the spin-charge coupling in the magnon self energy, between magnons and charge fluctuations in the partially-filled majority-spin band. Magnon decay into intermediate magnon states accompanied with charge fluctuations results in magnon-energy renormalization and magnon damping, as discussed below. 

It is straightforward to generalize to the case of localized spin-$S$ magnetic moments, 
corresponding to ${\cal N}=2S$ localized $\alpha$ orbitals. 
Replacing the fermion-magnon interaction vertex $J$ by $J\sqrt{2S}$, we obtain:
\begin{eqnarray}
\Sigma_{\rm total} ({\bf q},\omega) &=&
J^2 (2S) \sum_{\bf Q} \int \frac{d\Omega}{2\pi i} \; 
\frac{1}{\Omega+\omega_{\bf Q}^0 -i \eta}
\sum_{\bf k} \left [ \left ( \frac{1}
{\epsilon_{\bf k-q+Q}^{\uparrow +} -\epsilon_{\bf k}^{\uparrow -} +\omega -\Omega -i \eta} \right ) \right . \nonumber \\
& \times & \left . \frac{1}{2S} \left ( 1-\frac{2JS}
{\epsilon_{\bf k-q}^{\downarrow +} -\epsilon_{\bf k}^{\uparrow -} +\omega -i \eta} \right )^2 
\right ] .
\end{eqnarray}

In the strong-coupling (double-exchange) limit where $\epsilon_{\bf k} \ll 2JS$, the magnon self energy simplifies to:
\begin{eqnarray}
\Sigma_{\rm total} ({\bf q},\omega) &=&
J^2 (2S) \sum_{\bf Q} \sum_{\bf k} \left [ \left ( \frac{1}
{\epsilon_{\bf k-q+Q}^{\uparrow +} -\epsilon_{\bf k}^{\uparrow -} +\omega +\omega_{\bf Q} ^0 -i \eta} \right ) \frac{1}{2S} \left ( 
\frac{\epsilon_{\bf k-q} -\epsilon_{\bf k} +\omega }{2JS} \right )^2 
\right ] \nonumber \\
&=& \left ( \frac{1}{2S}\right )^2 
\sum_{\bf Q} \sum_{\bf k} 
\frac{(\epsilon_{\bf k-q} -\epsilon_{\bf k} +\omega )^2 }
{\epsilon_{\bf k-q+Q}^{\uparrow +} -\epsilon_{\bf k}^{\uparrow -} +\omega +\omega_{\bf Q} ^0 -i \eta} .
\end{eqnarray}
This magnon self energy has the following characteristic features:
i) it has the same energy order ($t$) as the bare magnon energy $\omega_{\bf q}^0$,
ii) it is not of the form $\sum_{\bf k} (\epsilon_{\bf k-q} -\epsilon_{\bf k})$,
implying deviation in the magnon dispersion from the Heisenberg form,
iii) it involves spin-charge coupling, resulting in magnon damping,
iv) it is down by the factor $1/2S$ relative to the classical magnon energy, 
and hence vanishes in the $S\rightarrow \infty$ limit.
Limiting behaviours of the resulting magnon dispersion --- non-Heisenberg form in the $J\rightarrow\infty$ limit and Heisenberg form in the $S\rightarrow \infty$ limit --- are in agreement with the exact calculations for the one-dimensional chain.\cite{kaplan_98}

The imaginary part of the strong-coupling magnon self energy:
\begin{equation}
\frac{1}{\pi} {\rm Im} \Sigma_{\rm total} ({\bf q},\omega) =
\left ( \frac{1}{2S}\right )^2 
\sum_{\bf Q} \sum_{\bf k} 
(\epsilon_{\bf k-q} -\epsilon_{\bf k} +\omega )^2 
\delta( \epsilon_{\bf k-q+Q}^{\uparrow +} -\epsilon_{\bf k}^{\uparrow -} + \omega +\omega_{\bf Q}^0 ) 
\end{equation}
yields finite magnon damping and linewidth at zero temperature, arising from magnon decay into intermediate magnon states accompanied with charge (majority-spin particle-hole) fluctuations. The above result for magnon damping due to the spin-charge coupling is in exact agreement with earlier strong-coupling results obtained using the Holstein-Primakoff transformation and $1/S$ expansion, but only after terms of third order in $1/S$ are included, amounting to a self-energy correction in the second-order magnon self energy.\cite{golosov_2000,shannon_2002}

\begin{figure}
\vspace*{-0mm}
\hspace*{-0mm}
\psfig{figure=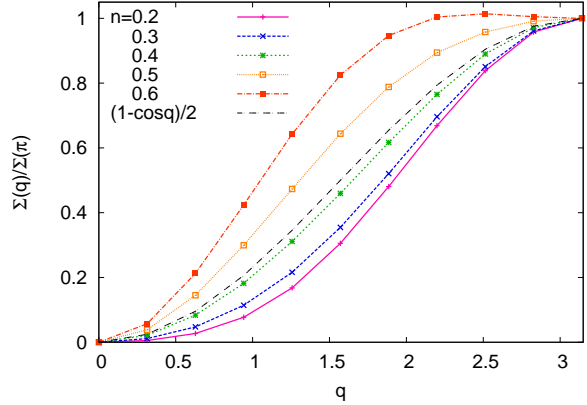,width=80mm}
\vspace{0mm}
\caption{The scaled self energy for the one-dimensional chain in the double-exchange limit, showing filling-dependent deviation from the Heisenberg form (dashed line), in agreement with the exact finite-size results of Ref. [12]. 
The deviation changes sign at $n\approx 0.43$.}
\label{avi1}
\end{figure}

\section{Results}

Fig. \ref{avi1} shows the self energy, calculated from Eq. 20 with $\omega=0$, for the one-dimensional chain in the double-exchange limit for $S=3/2$. The deviation in the magnon self energy from the Heisenberg form is seen to be filling dependent, with the deviation changing sign at $n \approx 0.43$, in agreement with the exact finite-size results.\cite{kaplan_98} 

\begin{figure}
\vspace*{-0mm}
\hspace*{-0mm}
\psfig{figure=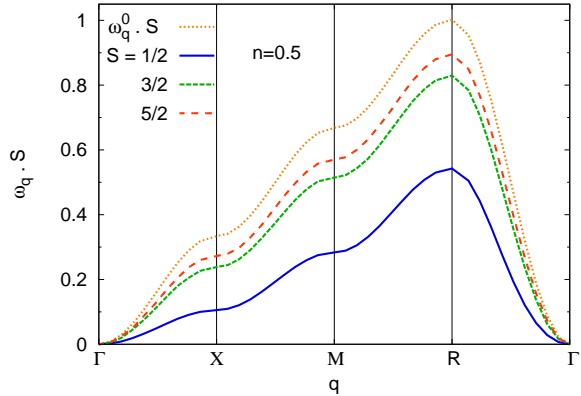,width=80mm}
\vspace{0mm}
\caption{The scaled renormalized magnon energy $\omega_{\bf q}.S$ for the sc lattice in the double-exchange limit, showing enhanced self-energy correction with decreasing spin quantum number $S$.}
\label{subrat}
\end{figure}

Fig. \ref{subrat} shows the scaled renormalized magnon dispersion $\omega_{\bf q}.S$ for the simple cubic lattice in the double-exchange limit. There is significant magnon-energy reduction due to spin-charge coupling, which is enhanced with decreasing spin quantum number $S$ highlighting the quantum $1/S$ effects, whereas it vanishes in the large $S$ limit. Even in the strong-coupling limit, the magnon dispersion shows deviation from the Heisenberg form, which is more pronounced at finite $J$ as discussed below.

\begin{figure}
\vspace*{-0mm}
\hspace*{-0mm}
\psfig{figure=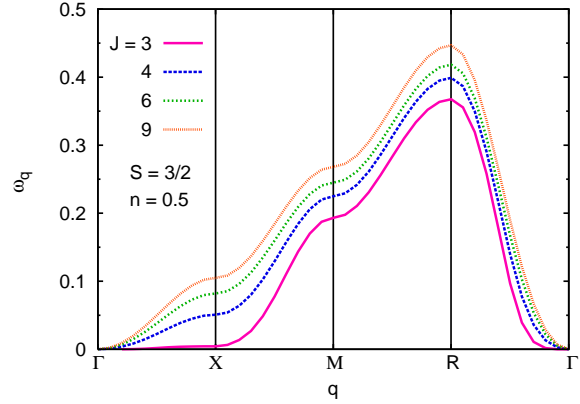,width=80mm}
\vspace{0mm}
\caption{The renormalized magnon dispersion for the simple cubic lattice in the intermediate-coupling regime, 
showing significant anomalous softening and deviation from the Heisenberg form.}
\label{dks}
\end{figure}

Fig. \ref{dks} shows the renormalized magnon dispersion for the simple cubic lattice in the intermediate-coupling regime. The anomalous momentum dependence of the magnon self energy results in significant anomalous softening which is robust even for $J \sim W$ where the Stoner gap is substantial. Even for large $J$, the ratio of magnon energies at X and R is noticeably smaller than 1/3. Also seen in the spin-wave excitation measurements of ferromagnetic manganites,\cite{Ye-PRB07} this feature indicates significant presence of second- ($J_2$) and third-neighbour ($J_3$) antiferromagnetic spin-spin interactions. 

\begin{figure}
\vspace*{-0mm}
\hspace*{-0mm}
\psfig{figure=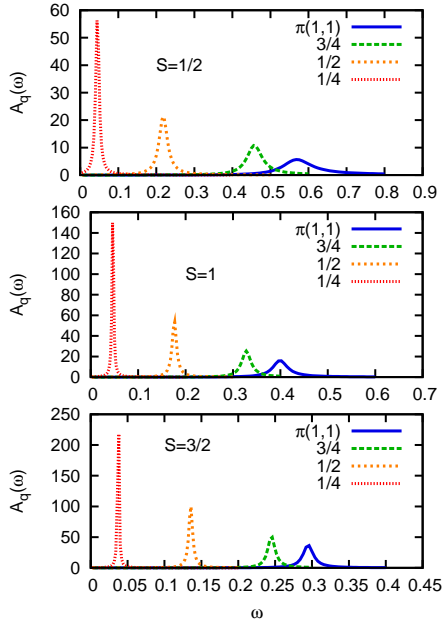,width=120mm}
\vspace{0mm}
\caption{The magnon spectral function for the square lattice in the double-exchange limit at filling n=0.82 and for different spin quantum numbers $S$, showing substantial broadening of zone-boundary modes due to magnon damping, especially for low $S$.}
\label{avi2}
\end{figure}

Fig. \ref{avi2} shows the behavior of the magnon spectral function for a square lattice 
from the zone center to the zone boundary along the $(1,1)$ direction
at band filling n=0.82 and in the double-exchange limit. 
Magnon damping, as indicated by the linewidth, is seen to be sharply enhanced for zone boundary modes, especially for low spin quantum number $S$ where quantum effects are significantly enhanced.



\begin{figure}
\vspace*{-0mm}
\hspace*{-0mm}
\psfig{figure=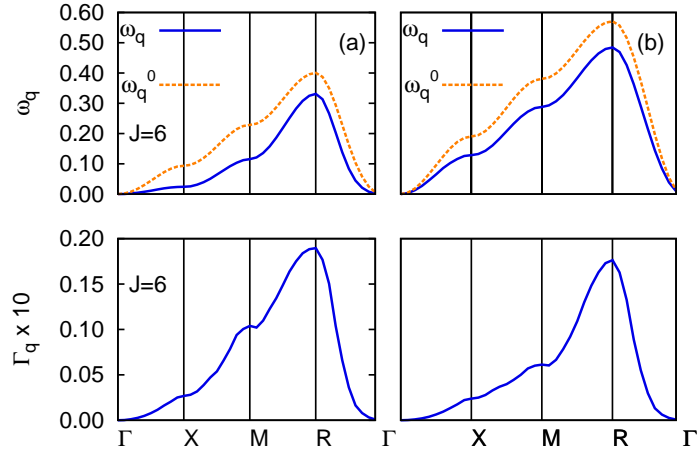,width=100mm}
\vspace{0mm}
\caption{Comparison of intermediate-coupling (a) and strong-coupling (b) results for the renormalized magnon dispersion 
$\omega_{\bf q}$ and magnon damping $\Gamma_{\bf q}$, for a simple cubic lattice with $n=0.7$ and $S=3/2$.}
\label{avi3}
\end{figure}

Fig. \ref{avi3} shows a comparison of intermediate-coupling and strong-coupling results for the renormalized magnon dispersion and magnon damping for a simple cubic lattice with $n=0.7$ and $S=3/2$. Here the magnon self energy was evaluated from Eq. (19) with $-\omega = \omega_{\bf q}^0$, the bare magnon energy. Relative to the magnon energy at R, magnon energies at X and M are substantially suppressed at finite $J$, highlighting the anomalous softening, while magnon damping is significantly enhanced near M. The strong momentum dependence of magnon damping, particularly along the M-R direction, highlights the role of charge fluctuations in the spin-charge coupling mechanism responsible for magnon damping. 

\begin{figure}
\vspace*{-0mm}
\hspace*{-0mm}
\psfig{figure=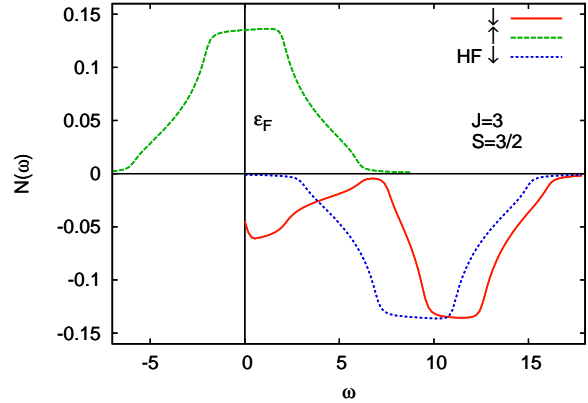,width=80mm}
\vspace{0mm}
\caption{The spin-resolved density of states for the simple cubic lattice at filling $n=0.5$, showing the transfer of minority-spin spectral weight and band narrowing due to correlation effects.}
\label{kbhaskar}
\end{figure}

\begin{figure}
\vspace*{-0mm}
\hspace*{-0mm}
\psfig{figure=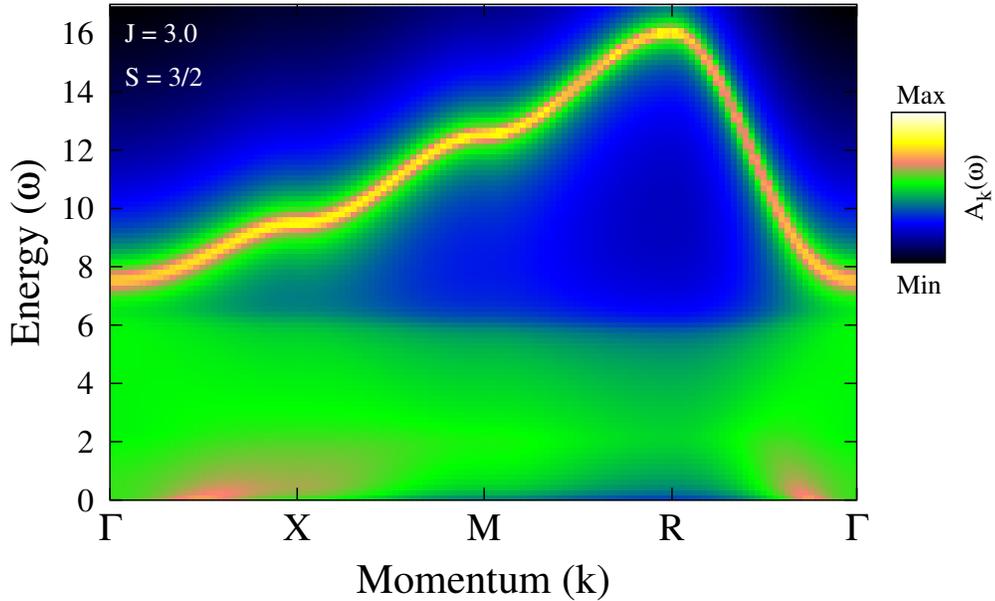,width=100mm,angle=-90}
\vspace{0mm}
\caption{Intensity plot of the minority-spin spectral function along symmetry directions, showing a nearly coherent high-energy branch with substantially reduced band width, and a strongly incoherent low-energy branch with hot spots between $\Gamma$-X and $\Gamma$-R.}
\label{gsap}
\end{figure}

Fig. \ref{kbhaskar} shows the spin-resolved density of states (DOS) of the band ($\beta)$ electrons on a simple cubic lattice at filling $n=0.5$. For spin-$\uparrow$, the bare (HF) DOS is exact as there are no self-energy corrections in the saturated ferromagnetic state at $T=0$. However, a spin-$\downarrow$ particle can decay into a magnon and a spin-$\uparrow$ particle due to the fermion-magnon interaction, as shown in the self-energy part of Fig. 2(a), resulting in considerable spin-$\downarrow$ spectral-weight transfer above the Fermi energy $\epsilon_{\rm F}$. Within an approximate resummation procedure which incorporates particle-particle correlations,\cite{hertz}
the spin-$\downarrow$ self energy for the $\beta$ orbital was calculated from:
\begin{equation}
\Sigma_{\beta\downarrow}({\bf q},\omega) = \frac{\Sigma_{\beta\downarrow}^{(0)}}{1-\Sigma^{(1)}_{\beta\downarrow}({\bf q},\omega)/\Sigma_{\beta\downarrow}^{(0)}} ,
\end{equation}
where $\Sigma_{\beta\downarrow}^{(0)} = 2\Delta_{\beta} = 2JS$ is the HF-level self energy,
and the first-order self energy incorporating electron-magnon interaction is given by:
\begin{eqnarray}
\Sigma_{\beta\downarrow}^{(1)} ({\bf q},\omega) &=& J^{2} (2S) \sum_{\bf Q}\int \frac{d\Omega}{2\pi i} \;
\chi^{+-}_{\rm RPA} ({\bf Q},\Omega)\left(\frac{1}{\omega-\Omega-\epsilon^{\uparrow +}_{{\bf k}-{\bf Q}} +i\eta}\right) 
\nonumber \\
& = & J^{2} (2S) \sum_{\bf Q} \frac{1}{\omega-\Omega^{0}_{\bf Q}-\epsilon^{\uparrow +}_{{\bf k}-{\bf Q}} +i\eta} .
\end{eqnarray}

Fig. \ref{gsap} depicts the energy-momentum dispersion of spin-$\downarrow$ electrons
for the above case in terms of an intensity plot (on a log scale) of the renormalized spectral function $A_{{\bf k}\downarrow}(\omega)$. The low-energy branch of the spectral-weight is strongly incoherent, with (relative) hot spots near the $\Gamma$ point between $\Gamma$-X and $\Gamma$-R. The high-energy branch remains nearly coherent, and the dispersion retains the bare character, although the renormalized band width is substantially reduced.

\section{Conclusions}
In conclusion, a purely fermionic representation for the FKLM was introduced in terms of a multi-orbital Hubbard model involving localized and band electrons representing local moments and itinerant fermions, and correlations effects were investigated using a systematic inverse-degeneracy diagrammatic scheme which explicitly preserves the continuous spin-rotation symmetry and hence the Goldstone mode. First-order quantum corrections to magnon excitations due to spin-charge coupling were investigated in the full range of interaction strength, and shown to result in strong magnon damping and anomalous softening for zone boundary modes, which accounts for several zone-boundary anomalies observed in recent spin-wave measurements of ferromagnetic manganites.

We find that in the strong-coupling limit, the magnon self energy for the one-dimensional chain exhibits filling-dependent deviation from the Heisenberg form, with the deviation changing sign near $n \sim 0.43$, in agreement with the exact finite-size results.\cite{kaplan_98} For the simple cubic lattice also, the renormalized magnon dispersion shows noticeable deviation from the Heisenberg form. Magnon decay into an intermediate-state magnon and majority-spin charge fluctuations due to spin-charge coupling lead to strong damping and line-broadening for zone-boundary modes in two and three dimensions, especially for low $S$. 

For finite $J$, spin-charge coupling is significantly enhanced, resulting in strong zone-boundary magnon damping and anomalous magnon softening even for large Stoner gap. This is in contrast to the RPA result\cite{wang_98} where both these features were found only in the weak-coupling regime where the Stoner gap is vanishingly small and the magnon branch merges with the Stoner continuun. However, in this weak-coupling regime, quantum corrections are strong enough to destabilize the ferromagnetic state. We also find that the ratio of magnon energies at X and R is noticeably smaller than 1/3, indicating significant presence of second $(J_2)$ and third-neighbour $(J_3)$ antiferromagnetic spin-spin interactions. This feature is also seen in a recent spin-wave excitation measurement of ferromagnetic manganites.\cite{Ye-PRB07} 

The interplay of electronic and magnetic excitations is highlighted by the fermion-magnon scattering of minority-spin electrons, resulting in substantial spectral-weight transfer to lower energies, with the additional spin-$\downarrow$ particle -- spin-$\uparrow$ hole process being responsible for the quantum exchange correction to spin stiffness. 
Intensity plot of the renormalized minority-spin spectral function showed a strongly incoherent low-energy branch, with (relative) hot spots near the $\Gamma$ point between $\Gamma$-X and $\Gamma$-R. The high-energy branch remained nearly coherent and the dispersion retained the bare character, although the renormalized band width is substantially reduced.

Finally, this purely fermionic representation and diagrammatic analysis not only provides insight into quantum effects on spin dynamics in FKLM systems such as manganites and DMS, but also allows seamless extention to include correlated band electrons, so that quantum fluctuation effects associated with both finite $S$ and electron correlation can be studied on an equal footing, which is particularly relevant for the ferromagnetic manganites.  


\end{document}